\input epsf.sty
\documentstyle[aps,twocolumn]{revtex}
\begin{document}
\draft
\title{Sub-millimeter tests of the gravitational inverse-square law: \\
a search for ``large'' extra dimensions}
\author{C.~D. Hoyle, U.~Schmidt, B.~R. Heckel, E.~G.~Adelberger, 
J.~H.~Gundlach, D.~J.~ Kapner, and H.~E.~Swanson}
\address{Department of Physics, University of Washington, 
Seattle, Washington 98195-1560}
\date{\today}
\maketitle
\begin{abstract}
Motivated by higher-dimensional theories that predict new
effects,
we tested the gravitational $1/r^2$ law at separations ranging 
down to 218 $\mu$m using a 
10-fold symmetric torsion pendulum and a
rotating 10-fold symmetric attractor. We improved previous
short-range constraints by up to a factor of 1000 and find no deviations from
Newtonian physics.
\end{abstract}
It is generally assumed that non-relativistic gravity obeys an 
inverse-square law for all distances greater than the Planck length 
$R_{\text {P}}=\sqrt{G \hbar /c^3}=1.6\times 10^{-33}$ cm. 
However, the gravitational interaction has 
only been tested\cite{kr:99,ho:85,mi:88,sm:00,su:94,ba:99,wi:93}
with good precision for separations greater than 1 cm and,
as summarized in Ref.~\cite{pr:99}, very little
is known about gravity at length scales below a few mm.
Recently theorists, using several different arguments, 
have suggested that the unexplored short-range regime of gravitation may hold 
profound surprises\cite{ar:98,ar:00,di:96,su:97}, {\em i.e.}
that the gravitational interaction could display fundamentally new behavior
in the mm regime. 

Many of these arguments are based on the notion, 
intrinisic to string or M
theory, of more than 3 spatial dimensions. To 
maintain consistency with a vast body of observations the extra 
dimensions must be ``curled up'' in very small regions, usually
assumed to be comparable to $R_{\text P}$, or else hidden in some other way
\cite{ra:99}.
It has recently
been noted\cite{ar:98,ar:00} that the enormous
discrepancy between natural mass scales of the Standard Model of
particle physics ($M_{\text {SM}}\approx 1$ TeV) 
and of gravity (the Planck mass 
$M_{\text P}=\sqrt{\hbar c/G}=1.2 \times 10^{16}$~TeV) 
could be eliminated if gravity propagates 
in {\em all} the space dimensions
while the other fundamental interactions are constrained to 
the three familiar dimensions. This unification scenario requires that
some of the
extra dimensions have radii $R^{\ast}$ that are 
large compared to $R_{\text P}$ with
\begin{equation}
R^{\ast}=\frac{\hbar c}{M^{\ast} c^2} \left(  \frac{M_{\text P}}{M^{\ast}}
\right)^{2/n}~,
\label{eq: R star}
\end{equation}
where $M^{\ast}$ is the unification scale (usually taken as $M_{\text {SM}}$)
and $n$ is the number of large extra dimensions.
The scenario with $n=1$
%a single large extra dimension 
is ruled out by astronomical data.
If there are 2 large extra dimensions, $R^{\ast}$  must be about 1 mm,
and the gravitational inverse-square law (which follows from
Gauss's Law in 3 spatial dimensions) will turn into a $1/r^4$-law
(Gauss's Law in 5 dimensions)
at distances much smaller than $R^{\ast}$.

Other theoretical considerations also suggest that new effects may show up
at short distances; string theories predict scalar particles 
(dilatons and moduli) that generate Yukawa
interactions which could be seen in tests of the $1/r^2$
law. If supersymmetry is broken at low energies these scalar particles would
produce mm-scale effects\cite{di:96,ka:00}.
Finally, there may be some significance to the observation\cite{su:97}
that the gravitational 
cosmological constant, $\Lambda \approx 3$~keV/cm$^3$, deduced from distant
Type 1A supernovae\cite{ri:98,pe:98} corresponds to a length scale 
$\sqrt[4]{\hbar c/\Lambda} \approx 0.1$~mm.
These, and other, considerations suggest that the Newtonian
gravitational potential should be replaced by a more general
expression\cite{ke:99} 
\begin{equation}
V(r)=-G \frac {m_1 m_2}{r}(1 + \alpha e^{-r/\lambda})~.
\label{eq: potential}
\end{equation}
%the dimensionless parameter $\alpha$ is predicted\cite{di:96,ke:99} 
%to lie between $10^5$ and $-1$ (
The simplest scenario with 2 large extra 
dimensions predicts $\lambda = R^{\ast}$ and 
$\alpha=3$ or $\alpha=4$ for compactification on
an 2-sphere or 2-torus, respectively\cite{ke:99}, while dilaton and moduli
exchange could produce forces with $\alpha$ as large as $10^5$ for
Yukawa ranges $\lambda \sim 0.1$~mm\cite{di:96}. 

This letter reports results of
a test of gravitational inverse-square law at length scales well below 1 mm. 
\vspace{0.16in}
Our instrument is shown in Fig.~\ref{fig: apparatus}.
%
%\vspace{0.1in}
\hspace*{0.5in}
\epsfysize=2.6in 
\epsfbox{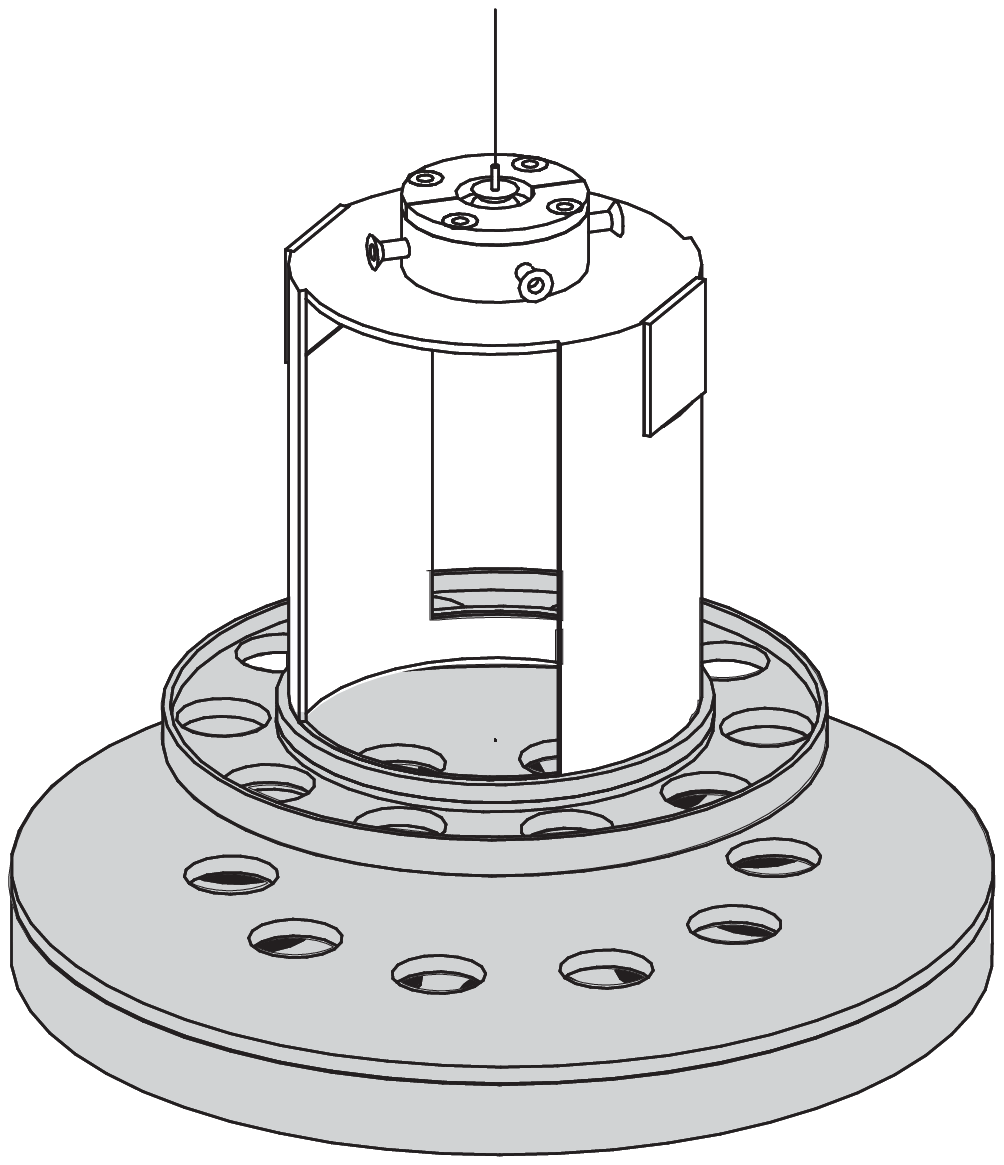}
\begin{figure}
\caption{Scale drawing of
the torsion pendulum and rotating attractor. The active components are shaded.
For clarity, we show
an unrealistically large 1.5 cm
vertical separation between pendulum and attractor, and
omit the BeCu membrane and the attractor
drive mechanism. The 4 horizontal screws were adjusted to
make the pendulum precisely level.}
\label{fig: apparatus}
\end{figure}
An 82~cm-long $20~\mu$m-diameter tungsten fiber, hanging from
an $x$-$y$-$z$-$\vartheta$ stage, supported
a torsion pendulum with 10-fold rotational symmetry. The pendulum was placed
above a 10-fold symmetric attractor that rotated slowly 
about the vertical axis of the pendulum. 
The active component of the pendulum
was a 2.002 mm thick aluminum annulus containing
ten equally-spaced 9.528 mm 
diameter holes, centered on a 55.33 mm diameter circle.
The active component of the attractor consisted 
of two coaxial 
copper disks.
The 1.846 mm thick upper 
disk had ten 9.535 mm diameter holes centered on a 55.31 mm
circle, while the 7.825 mm thick lower disk had ten 
12.690 mm diameter holes
centered on a 55.33 mm circle. The holes in the lower disk were 
rotated azimuthally by 18 degrees compared to the upper holes; this geometry
substantially reduced the signal from Newtonian gravity
but had little effect on a short-range signal.

We measured the torque on the pendulum, produced by the interactions between
the holes in the attractor and the pendulum 
for pendulum-to-attractor vertical
separations, $\zeta$, down to 218~$\mu$m.
Because of the 10-fold 
rotational symmetry of the instrument this torque varied
at 10 times the attractor rotation frequency $\omega$. Spurious torques
from electrical forces were minimized by placing a tightly
stretched, 11.4 cm diameter, 20 $\mu$m thick beryllium-copper membrane between
the attractor and pendulum, 48 $\mu$m above the upper surface of the 
attractor. The pendulum, mirrors, membrane and upper 
surface of the attractor were coated with gold, and
the pendulum and membrane were surrounded by a gold-plated copper enclosure.
Crucial dimensions of the pendulum and attractor were good to $\pm 2.5~\mu$m,
and the membrane, which had a mirror-like finish, was flat to within 
$\pm 15~\mu$m.

The torque on the pendulum was inferred from the 
pendulum twist angle, $\theta$, which we measured by reflecting an 
autocollimator light beam twice from either of 2 flat mirrors mounted 
on the pendulum.
The attractor rotation period, $\tau_{\text{A}}$, was set to $17 \tau_0$,
where $\tau_0 =2 \pi/\omega_0 \approx 399.6$~s was the period of 
pendulum's free torsional oscillations. 
%(We determined $\tau_0$ from each data point and found it to be
%constant to within xxx\% except for the 2 points at smallest $\zeta$ xxx.)
The attractor angle, $\phi=\omega t$, and
signals from the autocollimator, the ion-pump 
pressure gauge, 
a 2-axis electronic level on the instrument, and 8 temperature
sensors were digitized at
regular intervals, typically $\tau_{\text{D}}=\tau_0/36$, 
and stored in a small computer. 
Data, acquired continuously, were broken into ``cuts'' each of which
had exactly 2 oscillations of the $10 \omega$ signal.
For each ``cut'', the filtered\cite{filter} pendulum twist $\tilde{\theta}$
as a function of $\phi$ was fitted
with
\begin{equation}
\tilde{\theta} (\phi)=\sum_n \left[ b_n\sin n\phi + c_n\cos n\phi \right]
+ \sum_{m=0}^2
d_m P_m~,
\label{eq: fit}
\end{equation}
where the first sum ran over $n$=1, 3, 10, 20 and 30. The 
floating $n$=10, 20 and 30 terms yielded the fundamental and leading
harmonics of the torque signal; the small $n=1$ and $n=3$ terms were fixed
by fitting data covering one complete revolution of the attractor.
The $n$=3 coefficients accounted for the 3-fold symmetry of the attractor 
drive system and the $n$=1 coefficients, typically
consistent with zero, were used for diagnostic purposes. 
The $d_1$ and $d_2$ polynomial coefficients 
allowed for the slow unwinding of our
torsion fiber ($\lesssim \! 0.2\mu$r/h). 

Figure \ref{fig: data} shows a typical segment of fitted
autocollimator~data.\hfil\break
\vspace{-0.1in}
\hspace*{-0.32in}
\epsfxsize=4.0in
\epsfbox{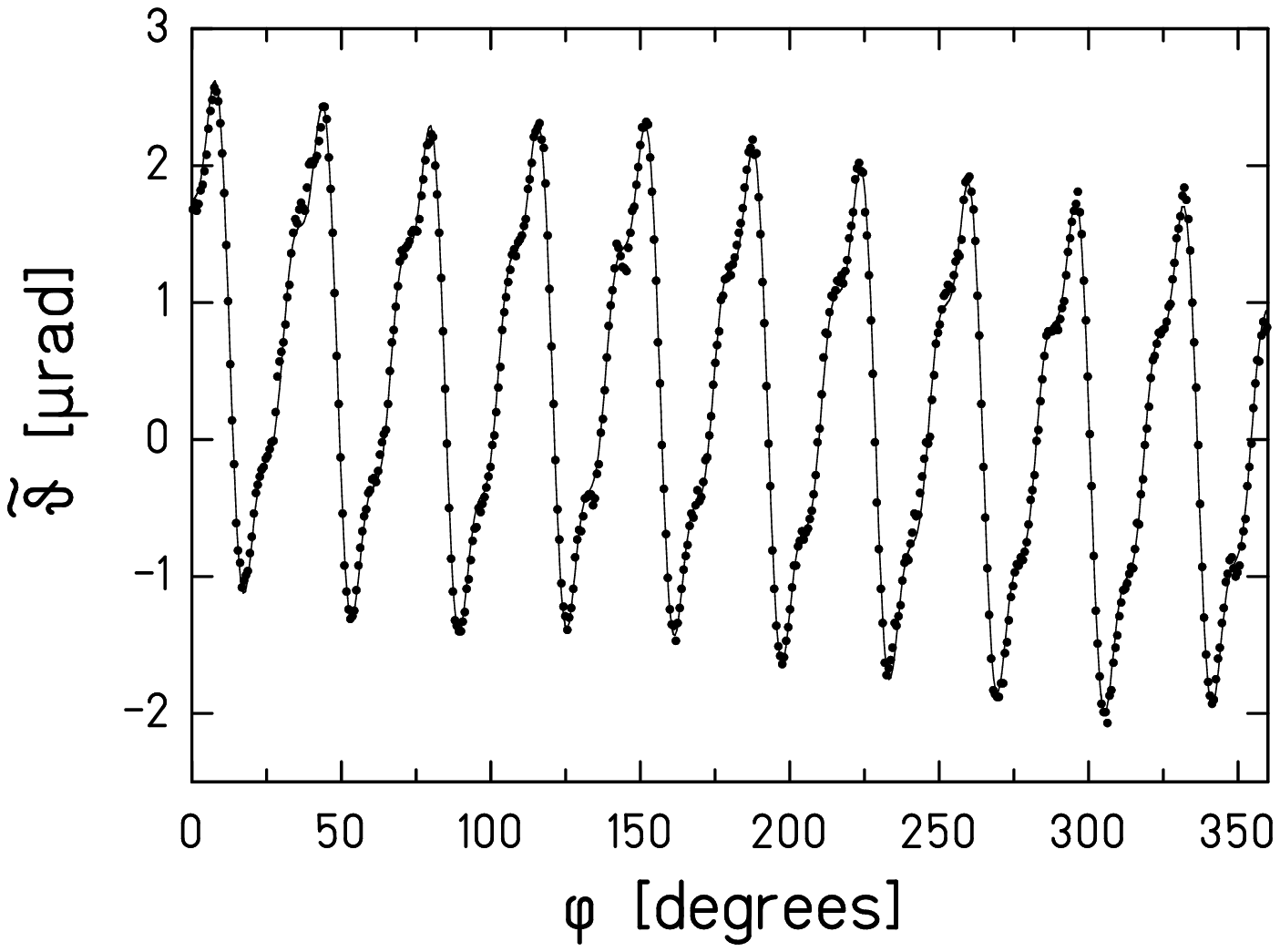}
\begin{figure}
\vspace{-0.1in}
\caption{Autocollimator data for one complete
revolution of the attractor, taken at $\zeta = 237~\mu$m. 
The curve is a fit to Eq.~\protect\ref{eq: fit}}
\label{fig: data}
\end{figure}
The resulting $n=10$, 20 and 30 torques are
\begin{equation}
T_n=I |(\tilde{b}_n+i\tilde{c}_n)(\omega_0^2-n^2\omega^2+2i\gamma n \omega)|
\sec(\frac{\pi n \omega}{2 \omega_0})~,
\end{equation}
where $I=128.4$ g-cm$^2$ is the calculated rotational inertia of the pendulum,
the superscript tildes on the $b$ and $c$ coefficients indicate corrections
for electronic time constants (see Ref.~\cite{sm:00}), and the damping 
coefficient $\gamma \approx 5 \times 10^{-6}$~s$^{-1}$ 
is the reciprocal of the observed decay time of the free torsional amplitude.
We measured $\omega_0$ for each data point and found it was constant to
within 0.3\%.

Our data sets consisted of the
$T_{10}$, $T_{20}$ and $T_{30}$ torques
%and $\tilde{b}_{30}$ coefficients 
as functions of separation.
Clearly, compared to Newtonian gravity, a short-range $1/r^2$-violating force 
will have both a
higher harmonic content and a stronger dependence on $\zeta$.
Data taken at $\zeta$'s
ranging from 218~$\mu$m to 10.78 mm and at several 
horizontal displacements $x$ and $y$, were compared to calculations
of the signals expected from the potential of
Eq.~\ref{eq: potential}.
% as well as from a $1/r^4$ force. 
The expected Newtonian (Yukawa) twists were computed to a 
precision better than 0.1(0.3$\alpha$)~nrad while a typical measured point 
had a statistical error
of 8 nrad. We found analytic solutions for 4
of the Newtonian integrals and 2 of the Yukawa integrals and evaluated the
remaining integrals numerically.
Figure~\ref{fig: b10} shows the $T_{10}$, $T_{20}$ and $T_{30}$
torques
together with the Newtonian predictions. We fitted these data,
for a series of $\lambda$ values, with
the physics parameter $\alpha$ and 4 constrained and 2 unconstrained
parameters describing the instrument. The constrained instrumental parameters
(the gap $\epsilon =  10 \pm 10~\mu$m 
between the two attractor disks, the ratio
of the ``masses'' of the holes in the
upper and lower disks
$\rho = 0.1329 \pm 0.0001$, 
a factor $f = 1.000 \pm 0.003$ representing the uncertainties in the torque 
scale and the ``mass'' of the pendulum holes, 
and $z_0 = 910 \pm 5~\mu$m the  $z$-micrometer reading at $\zeta=0$)
were determined by auxilliary measurements described below.
The unconstrained intrumental parameters $x_0$ and $y_0$ were the centered
values of $x$ and $y$ micrometers.
In all cases, the best-fit values of the constrained
experimental parameters agreed with their nominal values, 
the average and maximum deviations were 1.3$\sigma$ and 1.9$\sigma$,
respectively.
Our data are well described by Newtonian gravity with a $\chi^2$ per degree of
freedom of 1.03 with a probability of 41\%. The resulting
contraints on new physics of the form given in Eq.~\ref{eq: potential}
are shown in Fig.~\ref{fig: constraints}; scenarios with
\vspace*{-0.35in}
$\alpha \ge 3$ are excluded at 95\% confidence for $\lambda \ge 190~\mu$m.
\hspace*{-0.25in}
\epsfysize=4.288in
\epsfbox{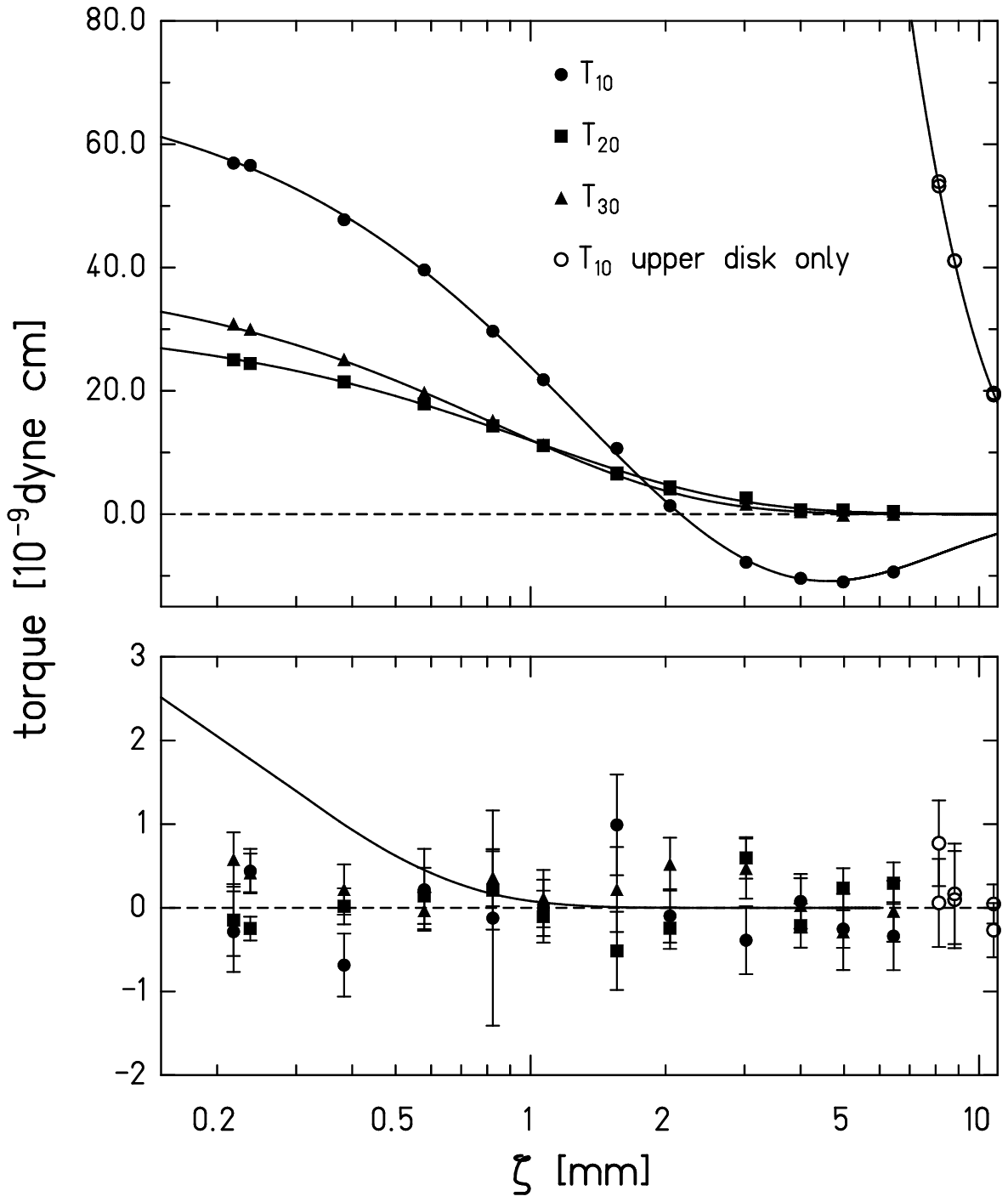}
\vspace*{-0.09in}
\begin{figure}
\vspace{-0.12in}
\caption{$T_{10}$, $T_{20}$ and $T_{30}$
torques, and the $10\omega$ torque from the upper attractor alone. Each point
contains at least 36 individual measurements; the $1\sigma$
error bars are derived from the scatter of the individual measurements. 
The upper panel shows the measured torques; the solid line
is the Newtonian prediction. The $T_{10}$ sign change at 
$z=2.2$~mm is due to cancellation from the bottom set of holes.
The lower panel displays the Newtonian fit
residuals; the solid curve shows the effect on $T_{10}$ of
an interaction 
with $\alpha=3$ and $\lambda=250~\mu$m.}
\label{fig: b10}
\end{figure}

Accurate alignment and control of the pendulum and attractor 
geometry and precise calibrations were essential in this experiment. 

The attractor geometry parameters were determined using
a measuring microscope, and the
``masses'' of the pendulum and attractor holes were computed from their
dimensions and the measured densities of the active components
of the pendulum and attractor.
The torque scale was calibrated gravitationally. Two 
small aluminum spheres, inserted into diametrically opposite
holes in a similar pendulum annulus, gave it a known
$q_{22}$ spherical multipole moment. A 
$Q_{22}$ attractor moment was created by two 
bronze spheres, centered on the torsion pendulum
and separated by 27.965 cm. These spheres were rotated 
around the pendulum axis by a turntable to produce a 
$(4.010\pm 0.001) \times
10^{-7}$ 
\vspace*{-0.45in}
dyne-cm $2\omega$ torque on the pendulum.
\hspace*{-0.32in}
\epsfysize=3.162in
\epsfbox{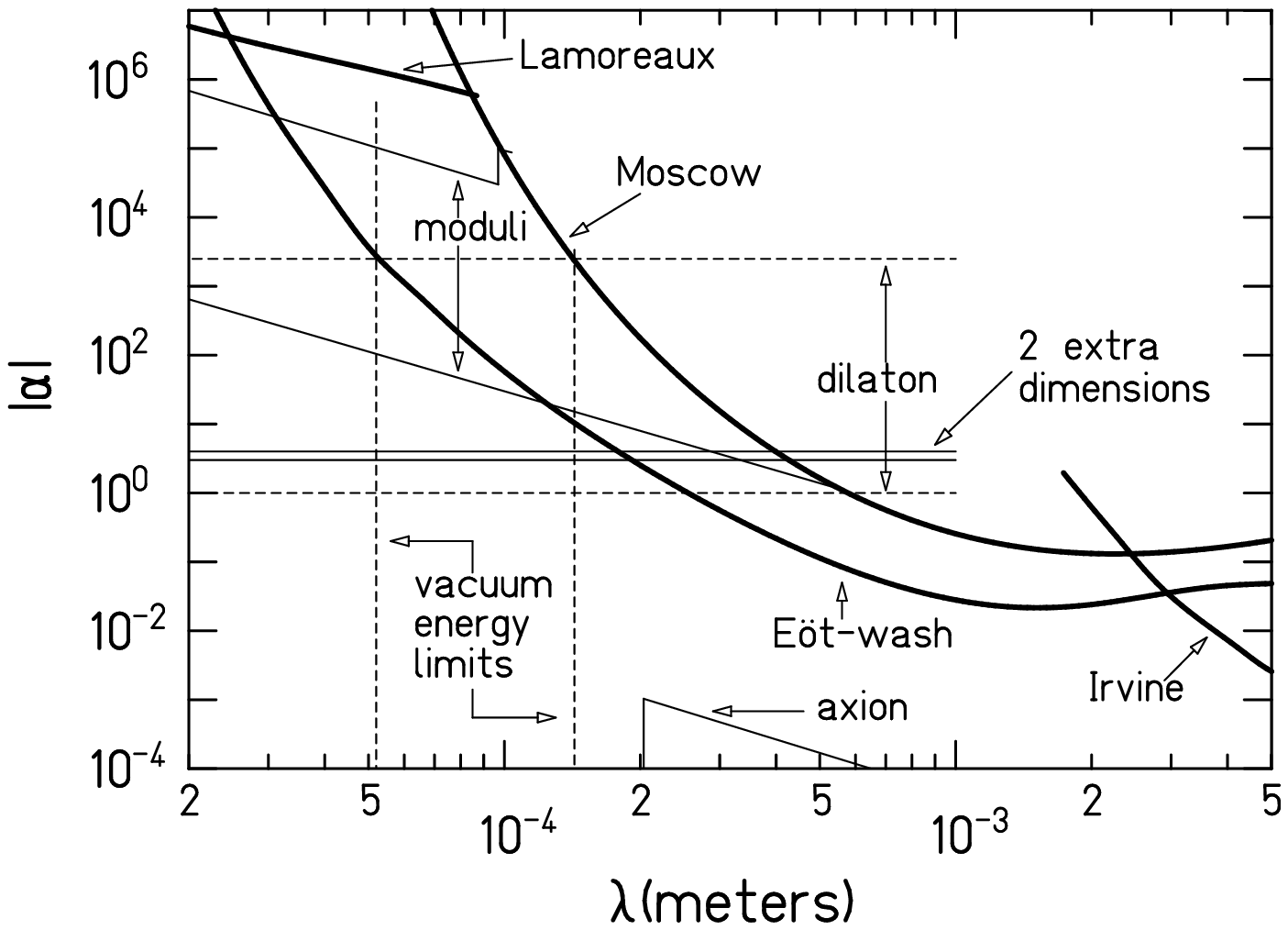}
\vspace{-0.3in}
\begin{figure}
%\vspace{-0.2in}
\caption{95\% confidence upper limits on $1/r^2$-law violating interactions
of the form given by Eq.~\protect\ref{eq: potential}. 
The region excluded by previous work\protect\cite{ho:85,mi:88,la:98} lies
above the heavy lines labeled Irvine, Moscow and Lamoreaux, respectively.
The data in Fig.~\protect\ref{fig: b10} imply the constraint
shown by the heavy line labeled E\"ot-Wash. 
Constraints from previous experiments and
the theoretical predictions are adapted from Ref.~\protect\cite{pr:99}, except 
for the dilaton prediction which is from Ref.~\protect\cite{ka:00}.}
\label{fig: constraints}
\end{figure}
The $z$-micrometer setting at vanishing vertical pendulum-to-membrane
separation was inferred from the capacitance between the 
pendulum and the membrane. (The membrane and the upper fiber attachment, 
which were normally grounded to the vacuum can, were
electrically isolated for these measurements.) We fitted
the capacitance as a function of the $z$-micrometer 
setting, see Fig.~\ref{fig: capacitor}, 
with an expression
for a plane electrode parallel to an infinite plate\cite{la:84}. 
(The function, which contained a leading-order expression 
for the fringing fields, was integrated over the pendulum's 
sinusoidal ``bounce'' motion.) 
The fixed vertical separation between the membrane and 
attractor ($48 \pm 3~\mu$m) and their parallelism (misalignment 
$\lesssim 0.2$ mrad) were measured using
depth micrometers and a special jig that could be substituted for the
membrane. A capacitive measurement gave a consistent value for this
separation.
The peak-to-peak vertical runout and ``wobble'' 
as the attractor rotated were $\lesssim 1.3~\mu$m and $\lesssim 5~\mu$m,
respectively.

The pendulum plane was adjusted to be 
horizontal to $< 0.2$~mrad by temporarily
replacing the normal membrane with 2 isolated semicircular copper disks that,
together with the pendulum, formed a differential capacitor. 
We then monitored the 
the $1\omega$ differential capacitance signal 
as $\vartheta$
%the upper fiber attachment 
was rotated at a steady rate and adjusted the pendulum's small trim screws
(shown in Fig.~\ref{fig: apparatus}) to minimize this signal. 
(The leveling precision was limited by the $\pm 74~\mu$m horizontal runout 
of the fiber as $\vartheta$ was rotated.)
The semicircular disks were removed and the reinstalled membrane was made 
parallel to the pendulum (to within $< 0.1$~mrad) by tilting the
entire instrument and finding the angle that gave the
smallest capacitance for a fixed value of $z$. 
Then the $x$ and $y$ micrometers were adjusted to recenter the pendulum 
on the 
\vspace*{-0.35in}
attractor by maximizing the $10\omega$, $20\omega$, and $30\omega$
signals.
\hspace*{-0.05in}
\epsfysize=2.49in
\epsfbox{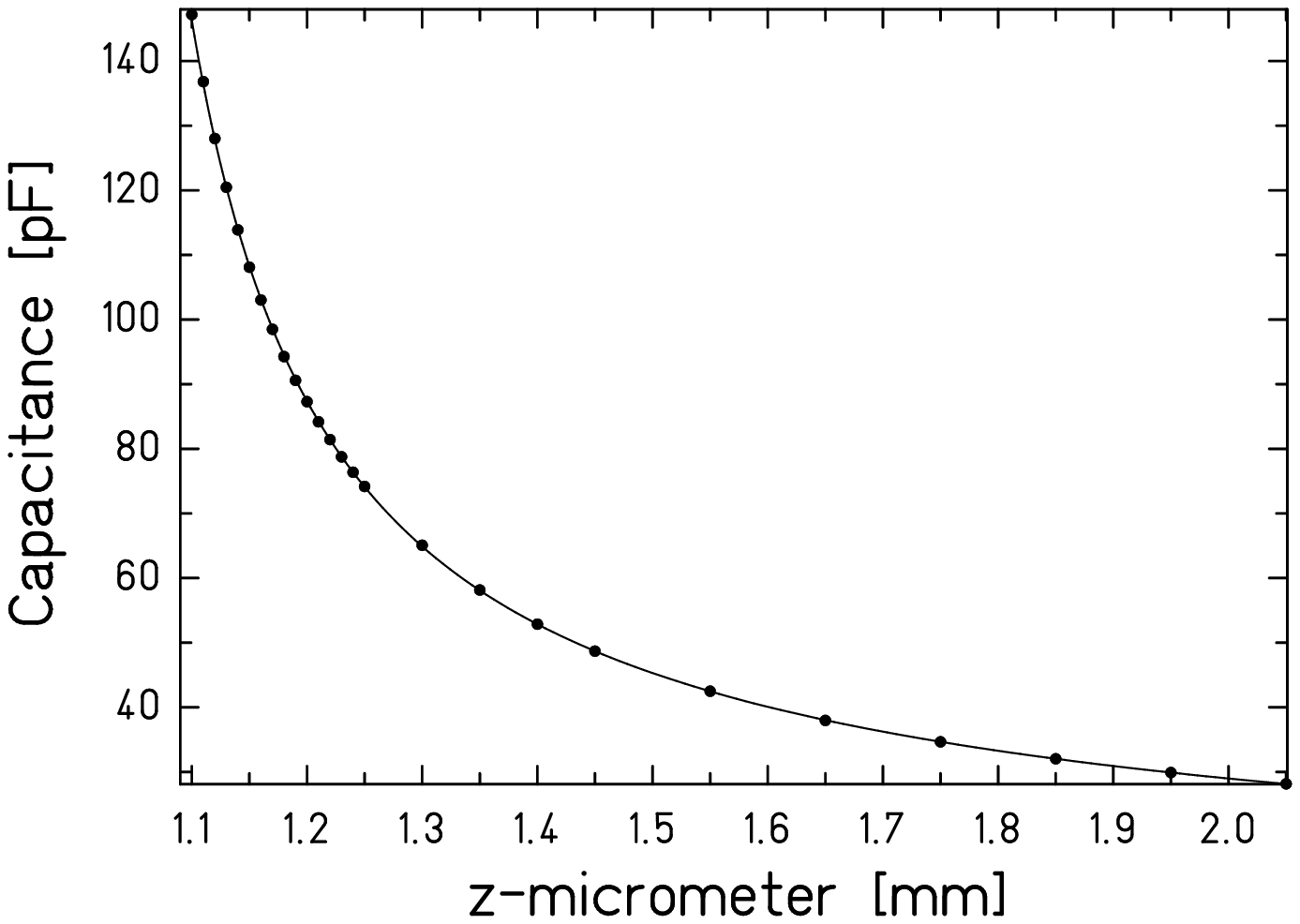}
\begin{figure}
\vspace*{-0.07in}
\caption{Capacitance measurements of the 
separation. The horizontal axis is the reading of the 
$z$-micrometer.
The fit indicated that the pendulum would touch the screen at 
$z=978 \pm 3~\mu$m.}
\label{fig: capacitor}
\end{figure}
We checked that likely non-gravitational
sources of systematic error had negligible effects on our results.
The attractor drive mechanism was largely nonmagnetic; the amplitude
of magnetic field variations at the site of the pendulum was $\leq 0.6~\mu$G
at the signal frequency. The measured
sensitivity of the pendulum to magnetic field
variations at the signal frequency was $57$ nrad/mG, corresponding
to a magnetic systematic error of 0.03 nrad.
The energy dissipated in the attractor motor warmed the attractor slightly.
A 5-fold increase in the 
motor's temperature rise, produced 
by attaching heat-dissipating resistors to the motor housing,
changed $\tilde{b}_{10}$ and $\tilde{b}_{20}$ by
$17 \pm 11$~nrad 
and $-4 \pm 9$~nrad 
respectively.
We found that an overall 98 mK change of the instrument's
temperature changed the equilibrium twist of the fiber by 1.17~$\mu$rad.
During normal operation, none of the sensors on the instrument had
temperature variations at the 10$\omega$ signal frequency that exceeded 
100~$\mu$K, indicating that spurious signals from temperature
variations were $\lesssim 1$~nrad. When the attractor rotation period was changed from
$7\tau_0$ to $43\tau_0$ (the normal period was $17\tau_0$),
$\tilde{b}_{10}$ and $\tilde{b}_{20}$
were unchanged to within
$5 \pm 15$~nrad and $11 \pm 31$~nrad,
respectively. 

Our results, interpreted in the simplest unification scenario
with 2 equal large extra dimensions, imply a unification scale 
given in Eq.~\ref{eq: R star} of $M^{\ast} \ge 3.5$~Tev.
We are now preparing a second-generation experiment with a different pendulum
and attractor that should provide higher precision and better sensitivity
at small $\lambda$.

We thank Profs. David Kaplan and Ann Nelson for helpful discussions, and 
Nathan Collins, Angela Kopp and Deb Spain for assistance with the experiment.
This work was supported primarily by the NSF (Grant PHY-9970987) and 
secondarily by the DOE.
\end{document}